\documentclass[seceq,preprint]{ptptex}
\usepackage{wrapft}

\usepackage{graphicx}

\pubinfo{Vol.~116, No.~4, October 2006}
\preprintnumber[3cm]{KUNS-2020}

\title{Gauge Theory and a Dirac Operator on a Noncommutative Space}

\author{Yoshinobu \textsc{Habara}}

\inst{Department of Physics, Kyoto University, Kyoto 606-8502, Japan}

\recdate{April 29, 2006}

\abst{As a tool to carry out the quantization of gauge theory on a noncommutative space, we present a Dirac operator that behaves as a line element of the canonical noncommutative space. Utilizing this operator, we construct the Dixmier trace, which is the regularized trace for infinite-dimensional matrices. We propose the possibility of solving the cosmological constant problem by applying our gauge theory on the noncommutative space.}

\newcommand{\tr}{\operatorname{tr}}
\notypesetlogo

\begin{document}
\maketitle

\section{Introduction}

Recently, noncommutative spaces whose space-time coordinates obey some algebraic relations, i.e. commutation relations, have been intensively investigated, because they are regarded as possible spaces within which to construct a theory of quantum gravity. The reason such spaces are considered in this role is that, although we cannot eliminate the ultraviolet divergence of quantum gravity in the ordinary way using renormalization, it is believed that the space-time noncommutativity $[\hat{x}^{\mu},\hat{x}^{\nu}]\! =\! i\theta^{\mu \nu}$ may lead to a relation of the form $\Delta x^{\mu}\Delta x^{\nu}\! \geq \! \theta^{\mu \nu}$, and thereby remove the infinitesimal region of space-time, i.e. the ultraviolet region, in the same manner as the Heisenberg commutation relation $[\hat{x},\hat{p}]\! =\! i\hbar$ leads to the Heisenberg uncertainty relation $\Delta x\Delta p\! \geq \! \hbar$ in the scheme of the first quantization. If indeed this is possible, we could obtain an ultraviolet divergence-free quantum gravity. In fact, it is natural to conjecture that operators $\hat{x}^{\mu}$ which satisfy commutation relations of the above form would be represented by matrices and have discrete spectra. However, as is obvious if we take the trace of both sides of the above commutation relation, we have the problem that the operators $\hat{x}^{\mu}$ are represented by infinite-dimensional matrices, and the naive trace, constituting the volume integration of the space-time, gives infinity. For this reason, to this time, we have studied noncommutative spaces that can be expressed by finite-dimensional matrices, such as noncommutative spherical surfaces (see Refs.~\citen{rf:watamura,rf:klimcik}, etc.). Or, to treat the canonical noncommutative space $[\hat{x}^{\mu},\hat{x}^{\nu}]\! =\! i\theta^{\mu \nu}$, the most well-known method is that utilizing the $\ast$-product, which shifts the noncommutativity of the operators $\hat{x}^{\mu}$ to the product of the ordinary commutative coordinates $x^{\mu}$ (Ref.~\citen{rf:seiberg}, etc.). However, this results in other problems, such as UV/IR mixing, the appearance of an infinite number of derivatives and unitality. In order to avoid such problems, do not employ this method here.

In this paper, we take the point of view that the Heisenberg algebra is the very procedure to make the phase space noncommutative, and the algebra of the noncommutative space is an extension of the Heisenberg algebra, as discussed in Refs.~\citen{rf:y1} and \citen{rf:y2}. Of course, since the operators $\hat{x}^{\mu}$ which generate the noncommutative space are infinite-dimensional matrices and their naive traces are infinite, we must define a regularized trace, known as a Dixmier trace,\cite{rf:connes,rf:landi} which accompanies a Dirac operator that acts as a regulator. In this way, we obtain a finite trace. As the basis of the representation of $\hat{x}^{\mu}$, we employ that of the harmonic oscillator, which is very familiar as an infinite-dimensional representation. Then we can regard our Dirac operator as the noncommutative extension of the ordinary Dirac operator on the commutative space and interpret its inverse as the line element of the noncommutative space.

The organization of this paper is as follows. In \S 2, we review the introduction of a gauge theory on a canonical noncommutative space presented in Refs.~\citen{rf:y1} and \citen{rf:y2}. In \S 3, we define the Dirac operator and a regularized trace for infinite-dimensional matrices called the Dixmier trace on the basis of the eigenfunctions of the harmonic oscillator, and we construct the complete system to expand the fields. Then we carry out the volume integration of the action of our gauge theory. In \S 4, we present our conclusion and propose a possible resolution of the cosmological constant problem.

\section{Noncommutative space and the action of the gauge field}

In this section, we introduce a canonical noncommutative space as an algebra that consists of some generators and review the construction of gauge fields which belong to it and their action, following Refs.~\citen{rf:y1} and \citen{rf:y2}. Here, although we take the space-time dimensionality to be $d\! =\! 4$, it may be trivial to extend it to the general even-dimensional case $d\! =\! 2r\> (r\! =\! 1,2,\cdots )$. The signature of the space-time metric can be either Minkowskian, $\eta^{\mu \nu}\! =\! \text{diag}(-1,1,1,1)$, or Euclidean, $\eta^{\mu \nu}\! =\! \text{diag}(1,1,1,1)$. However, as we see in \S 3, the geometric properties, e.g., the meaning of the volume integration, are clearer in the Euclidean case. In the following, we set $\hbar \! =\! c\! =1$.

First, we introduce the algebra $\mathcal{A}\oplus End_{\mathcal{A}}\mathcal{H}$ generated by the algebraic structure 
\begin{align}
	& \left\{ \begin{array}{l}
	[\hat{x}^{\mu},\hat{x}^{\nu}]=+i\theta^{\mu \nu}, \\{}
	[\hat{x}^{\mu},\hat{p}_{\nu}]=i\delta^{\mu}_{\> \nu}, \\{}
	[\hat{p}_{\mu},\hat{p}_{\nu}]=-i\theta_{\mu \nu}^{-1},
	\end{array} \right. \quad 
	\left\{ \begin{array}{l}
	[\hat{x}^{\prime \mu},\hat{x}^{\prime \nu}]=-i\theta^{\mu \nu}, \\{}
	[\hat{x}^{\prime \mu},\hat{p}_{\nu}^{\prime}]
	=i\delta^{\mu}_{\> \nu}, \\{}
	[\hat{p}_{\mu}^{\prime},\hat{p}_{\nu}^{\prime}]
	=+i\theta_{\mu \nu}^{-1},
	\end{array} \right. \nonumber \\
	& \quad [\hat{x}^{\mu},\hat{x}^{\prime \nu}]
	=[\hat{x}^{\mu},\hat{p}_{\nu}^{\prime}]
	=[\hat{p}_{\mu},\hat{p}_{\nu}^{\prime}]=0, \nonumber \\
	& \quad \text{with } \mu ,\nu =0,1,2,3, 
	\label{eq:1}
\end{align}
which characterizes a canonical noncommutative space. Here, $\theta^{\mu \nu}$ expresses the space-time noncommutativity and is a non-dynamical constant value which satisfies $\theta^{\mu \nu}\! =\! -\theta^{\nu \mu}$. $\mathcal{H}$ is a Hilbert space consisting of the square-integrable functions on which the operators $\hat{x}^{\mu}, \hat{p}_{\mu} \in \mathcal{A}$ and their endomorphisms $\hat{x}^{\prime \mu}, \hat{p}_{\mu}^{\prime} \in End_{\mathcal{A}}\mathcal{H}$ act, and below we find that it coincides with the Fock space of a harmonic oscillator system.  We regard this noncommutative space as an extension of the Heisenberg algebra, which is the starting point of the first quantization. That is, we would like to think of the Heisenberg algebra as a kind of noncommutative space whose coordinates $x^{\mu}$ do not commute with the momenta $p_{\mu}$, and then we impose noncommutativity among the $x^{\mu}$ themselves and among the $p_{\mu}$ themselves. We assume that (or use dimensional analysis to argue that) $\theta^{\mu \nu}$ is of the order of the square of the Planck scale: $\theta^{\mu \nu}\! \sim \! m_p^{-2}$. This is consistent with the general conjecture that space-time noncommutativity may emerge at the Planck scale, $m_p\! \sim \! l_p^{-1}$. In the following, because the anti-symmetric matrix $\theta^{\mu \nu}$ can be normalized by a linear transformation, we set 
\begin{align}
	\theta^{\mu \nu}
	=\theta \left( \begin{array}{cccc}
	0 & 1 & 0 & 0 \\
	-1 & 0 & 0 & 0 \\
	0 & 0 & 0 & 1 \\
	0 & 0 & -1 & 0 
	\end{array} \right), 
	\label{eq:2}
\end{align}
for simplicity. Thus our gauge theory on $\mathcal{A}\oplus End_{\mathcal{A}}\mathcal{H}$ contains only {\it one} parameter $\theta$ and includes the space-time and momentum noncommutativities that do not contribute to the field theory defined at an energy scale below the Planck scale (i.e., at a spatial scale above the Planck length).

The generators (\ref{eq:1}) of the algebra $\mathcal{A}\oplus End_{\mathcal{A}}\mathcal{H}$ are obtained as 
\begin{align}
	& \left\{ \begin{array}{l}
	\hat{x}^{\mu}=\frac{1}{2}x^{\mu}+i\theta^{\mu \nu}\partial_{\nu} \\
	\hat{p}_{\mu}=-i\partial_{\mu}-\frac{1}{2}\theta_{\mu \nu}^{-1}x^{\nu}
	\end{array} \right. , \quad 
	\hat{p}_{\mu}=-\theta_{\mu \nu}^{-1}\hat{x}^{\nu}, \nonumber \\
	& \left\{ \begin{array}{l}
	\hat{x}^{\prime \mu}=\frac{1}{2}x^{\mu}-i\theta^{\mu \nu}
	\partial_{\nu} \\
	\hat{p}_{\mu}^{\prime}=-i\partial_{\mu}+\frac{1}{2}
	\theta_{\mu \nu}^{-1}x^{\nu}
	\end{array} \right. , \quad 
	\hat{p}_{\mu}^{\prime}=\theta_{\mu \nu}^{-1}\hat{x}^{\prime \nu}, 
	\label{eq:3}
\end{align}
by using the operators that act on the Hilbert space $\mathcal{H}$ of the square-integrable functions. Just as in, for example, the case of the 4-dimensional harmonic oscillator, which is generated by eight operators, i.e. four creation operators, $a_{\mu}^{\dagger}\! =\! \frac{1}{\sqrt{2}}(x_{\mu}\! -\! \partial_{\mu})$, and four annihilation operators, $a_{\mu}\! =\! \frac{1}{\sqrt{2}}(x_{\mu}\! +\! \partial_{\mu})$, the eight generators $\hat{x}^{\mu},\> \hat{x}^{\prime \mu}\> (\mu \! =\! 0,\cdots ,3)$ of $\mathcal{A}\oplus End_{\mathcal{A}}\mathcal{H}$ are constructed as linear combination of the eight operators $x^{\mu}$ and $-i\partial_{\mu}$.

Next, let us introduce gauge fields $A_{\mu}$ and connections $\nabla_{\mu}$ in the natural way.\cite{rf:y1,rf:y2} In analogy to the connections on the ordinary commutative space-time, defined as $\nabla_{\mu}=\partial_{\mu}+iA_{\mu}(x)$, we define 
\begin{align}
	& \nabla_{\mu}\equiv i\hat{p}_{\mu}^{\prime}-i\hat{A}_{\mu}, 
	\nonumber \\
	& \nabla_{\mu}^{\prime}\equiv i\hat{p}_{\mu}+i\hat{A}_{\mu}^{\prime} 
	\label{eq:4}
\end{align}
on $\mathcal{A}\oplus End_{\mathcal{A}}\mathcal{H}$. For all $a\in \mathcal{A},\> a^{\prime}\in End_{\mathcal{A}}\mathcal{H}$ and $f\in \mathcal{H}$, if we impose the Leibnitz rule 
\begin{align*}
	& \nabla_{\mu}(a^{\prime}f)=[i\hat{p}_{\mu}^{\prime},a^{\prime}]f
	+a^{\prime}\nabla_{\mu}f, \nonumber \\
	& \nabla_{\mu}^{\prime}(af)=[i\hat{p}_{\mu},a]f
	+a\nabla_{\mu}^{\prime}f,
\end{align*}
the gauge fields must satisfy the conditions 
\begin{align}
	& [\hat{A}_{\mu},a^{\prime}]=0 \> \Longrightarrow \> 
	\hat{A}_{\mu}\in \mathcal{A}, \nonumber \\
	& [\hat{A}_{\mu}^{\prime},a]=0 \> \Longrightarrow \> 
	\hat{A}_{\mu}^{\prime}\in End_{\mathcal{A}}\mathcal{H}.
	\label{eq:5}
\end{align}
Therefore, (\ref{eq:4}) becomes 
\begin{align}
	& \nabla_{\mu}=i\hat{p}_{\mu}^{\prime}-i\hat{A}_{\mu}(\hat{x}), 
	\nonumber \\
	& \nabla_{\mu}^{\prime}=i\hat{p}_{\mu}
	+i\hat{A}_{\mu}(\hat{x}^{\prime}).
	\label{eq:6}
\end{align}
For these gauge fields, the gauge transformations are realized in the algebras $\mathcal{A}$ and $End_{\mathcal{A}}\mathcal{H}$, respectively. That is, for unitary operators $U(\hat{x})\in \mathcal{A}$ and $U(\hat{x}^{\prime})\in End_{\mathcal{A}}\mathcal{H}$, the connections $\nabla_{\mu}$ and $\nabla_{\mu}^{\prime}$ and the gauge fields $A_{\mu}(\hat{x})$ and $A_{\mu}(\hat{x}^{\prime})$ transform as 
\begin{align}
	& \nabla_{\mu}\to U^{\dagger}(\hat{x})\nabla_{\mu}U(\hat{x}) 
	\quad \Rightarrow \quad 
	A_{\mu}(\hat{x})\to U^{\dagger}(\hat{x})A_{\mu}(\hat{x})U(\hat{x}), 
	\nonumber \\
	& \nabla_{\mu}^{\prime}\to U^{\dagger}(\hat{x}^{\prime})
	\nabla_{\mu}^{\prime}U(\hat{x}^{\prime}) 
	\quad \Rightarrow \quad 
	A_{\mu}(\hat{x}^{\prime})\to U^{\dagger}(\hat{x}^{\prime})
	A_{\mu}(\hat{x}^{\prime})U(\hat{x}^{\prime}).
	\label{eq:7}
\end{align}
From the connections (\ref{eq:6}), we can construct field strengths which belong to $\mathcal{A}$ and $End_{\mathcal{A}}\mathcal{H}$, respectively: 
\begin{align}
	& \hat{F}_{\mu \nu}\equiv [\nabla_{\mu},\nabla_{\nu}]
	=-[A_{\mu}(\hat{x}),A_{\nu}(\hat{x})]-i\theta_{\mu \nu}^{-1}, 
	\nonumber \\
	& \hat{F}_{\mu \nu}^{\prime}
	\equiv [\nabla_{\mu}^{\prime},\nabla_{\nu}^{\prime}]
	=-[A_{\mu}(\hat{x}^{\prime}),A_{\nu}(\hat{x}^{\prime})]
	+i\theta_{\mu \nu}^{-1}. 
	\label{eq:8}
\end{align}
These field strengths are transformed as 
\begin{align}
	& \hat{F}_{\mu \nu}\to U^{\dagger}(\hat{x})\hat{F}_{\mu \nu}
	U(\hat{x}), \nonumber \\
	& \hat{F}_{\mu \nu}^{\prime}\to U^{\dagger}(\hat{x}^{\prime})
	\hat{F}_{\mu \nu}^{\prime}U(\hat{x}^{\prime}) 
	\label{eq:9}
\end{align}
under the gauge transformation (\ref{eq:7}).

Lastly, the Yang-Mills action on the noncommutative space $\mathcal{A}\oplus End_{\mathcal{A}}\mathcal{H}$ that is gauge invariant is obtained in the form 
\begin{align}
	S_{\text{YM}} 
	& =\frac{1}{4}\int \hat{F}_{\mu \nu}\hat{F}^{\mu \nu}
	+\frac{1}{4}\int \hat{F}_{\mu \nu}^{\prime}\hat{F}^{\prime \mu \nu} 
	\nonumber \\
	& =\frac{1}{4}\int [A_{\mu}(\hat{x}),A_{\nu}(\hat{x})]
	[A^{\mu}(\hat{x}),A^{\nu}(\hat{x})]
	+\frac{1}{4}\int [A_{\mu}(\hat{x}^{\prime}),A_{\nu}(\hat{x}^{\prime})]
	[A^{\mu}(\hat{x}^{\prime}),A^{\nu}(\hat{x}^{\prime})] 
	\nonumber \\
	& \qquad \qquad -\frac{1}{2}\theta_{\mu \nu}^{-1}\theta^{-1 \mu \nu}
	\int 1, 
	\label{eq:10}
\end{align}
utilizing an integral $\int$ defined in the next section. Here, the gauge fields $A_{\mu}(\hat{x})$ and $A_{\mu}(\hat{x}^{\prime})$ are different only in their arguments and are essentially the same fields; i.e. they have the same coefficients when they are expanded in terms of $\hat{x}^{\mu}$ and $\hat{x}^{\prime \mu}$, respectively. For the experimental energy scale we consider $\theta \! \to \! 0$, we have $\hat{x}^{\mu}=\hat{x}^{\prime \mu}$, and in this case, (\ref{eq:10}) reduces to the action of the ordinary Yang-Mills theory. Also, note that this understanding is natural when we consider the number of the degrees of freedom of the gauge field.

We should note that, although the integral $\int$ is in principle the trace of matrices, these integrals of $\hat{x}^{\mu}$, etc., usually diverge and do not satisfy the cyclic property $\int ab=\int ba$, since they are infinite-dimensional matrices, as is obvious from the algebra (\ref{eq:1}). That is, the action $S_{\text{YM}}$ is not invariant naively under the gauge transformation (\ref{eq:7}). For this reason, in the next section, we define this integral explicitly by utilizing a special trace known as the Dixmier trace, and find how the cyclic property necessary for the gauge invariance is realized.

A comment on the gauge transformation (\ref{eq:7}) is also in order. This transformation is not exactly the same as the ordinary gauge transformation of Yang-Mills theory but, rather, it is a unitary transformation on the algebra $\mathcal{A}\oplus End_{\mathcal{A}}\mathcal{H}$, which represents the noncommutative space. In other words, it corresponds to a geometric transformation of the coordinates of the noncommutative space. The so-called $U(1)$ gauge transformation is included in (\ref{eq:7}) as $U(\hat{x})\! =\! e^{i\alpha}\cdot {\bf 1}$ for global $U(1)$ and $U(\hat{x})\! =\! e^{i\gamma (\hat{x})}$ for local $U(1)$ (and also for $\hat{x}^{\prime}$), where $\alpha$ is a scalar, ${\bf 1}$ is the unit operator, and $\gamma (\hat{x})\! =\! \gamma^{\dagger}(\hat{x})$ in $\mathcal{A}\oplus End_{\mathcal{A}}\mathcal{H}$. Then, if we want to extend it to the general gauge group, for example, the $SU(N)$ gauge group, it is sufficient to alter the noncommutative space to $(\mathcal{A}\oplus End_{\mathcal{A}}\mathcal{H})\otimes SU(N)$. However, we employ $U(1)$ in this article, because this extension is not essential for the analysis given here.

\section{The Dirac operator and the volume integration}

Now we construct the Dirac operator for the canonical noncommutative space (\ref{eq:1}) and define the integral $\int$ that produces finite values for the traces of the infinite-dimensional matrices. (See Refs.~\citen{rf:connes} and \citen{rf:landi} for the formulation of the general theory of the Dirac operator and the Dixmier trace, of which we employ a special case in this section.)

To begin, we consider the following linear combinations of the generators $\hat{x}^{\mu}$ and $\hat{x}^{\prime \mu}$ of the noncommutative space: 
\begin{align}
	& a_1=\frac{1}{\sqrt{2\theta}}(\hat{x}^0+i\hat{x}^1), \quad 
	a_1^{\dagger}=\frac{1}{\sqrt{2\theta}}(\hat{x}^0-i\hat{x}^1), 
	\nonumber \\
	& a_2=\frac{1}{\sqrt{2\theta}}(\hat{x}^2+i\hat{x}^3), \quad 
	a_2^{\dagger}=\frac{1}{\sqrt{2\theta}}(\hat{x}^2-i\hat{x}^3), 
	\nonumber \\ 
	& a_1^{\prime}=\frac{1}{\sqrt{2\theta}}(\hat{x}^{\prime 0}
	-i\hat{x}^{\prime 1}), \quad 
	a_1^{\prime \dagger}=\frac{1}{\sqrt{2\theta}}(\hat{x}^{\prime 0}
	+i\hat{x}^{\prime 1}), 
	\nonumber \\
	& a_2^{\prime}=\frac{1}{\sqrt{2\theta}}(\hat{x}^{\prime 2}
	-i\hat{x}^{\prime 3}), \quad 
	a_2^{\prime \dagger}=\frac{1}{\sqrt{2\theta}}(\hat{x}^{\prime 2}
	+i\hat{x}^{\prime 3}). 
	\label{eq:11}
\end{align}
These new operators form the algebra of the 4-dimensional harmonic oscillator: 
\begin{align}
	& [a_i,a_j^{\dagger}]=[a_i^{\prime},a_j^{\prime \dagger}]=\delta_{ij}, 
	\quad (i,j=1,2) \nonumber \\
	& \text{otherwise}=0.
	\label{eq:12}
\end{align}
Then, as stated above, the Hilbert space $\mathcal{H}$ on which the generators $\hat{x}^{\mu}$ and $\hat{x}^{\prime \mu}$ act is that of the square-integrable eigenfunctions of the harmonic oscillator. However, from the method of construction of the gauge field and its action (\ref{eq:10}), and considering (\ref{eq:11}), we find that the contributions of $\hat{x}^{\mu}\in \mathcal{A}$ and $\hat{x}^{\prime \mu}\in End_{\mathcal{A}}\mathcal{H}$ are completely independent, and therefore we should divide the 4-dimensional harmonic oscillator into two parts, thereby obtaining a system of 2+2 dimensions. Below we find that this separation is correct from the point of view of the space-time dimensionality. Therefore we first study the 2-dimensional harmonic oscillator, consisting of $a_i$ and $a_i^{\dagger}\> (i\! =\! i,2)$. In general, it is obvious that we treat an $r$-dimensional harmonic oscillator when the space-time dimensionality is $d\! =\! 2r$.

We define the Dirac operator $D$ that describes an infinitesimal length on noncommutative space and plays the role of regulating the trace of the infinite-dimensional matrices as 
\begin{align}
	D^2 & \equiv \frac{1}{\theta}(a_1^{\dagger}a_1+a_2^{\dagger}a_2+1) 
	\nonumber \\
	& =\frac{1}{2}\left\{(\hat{p}^0)^2+(\hat{p}^1)^2+(\hat{p}^2)^2
	+(\hat{p}^3)^2\right\}, \nonumber \\
	& \! \! \! \! \text{eigenvalue: }\mu_N(D^2)=\frac{1}{\theta}(N+1), 
	\nonumber \\
	& \! \! \! \! \text{degeneracy: }m_N=N+1.
	\label{eq:13}
\end{align}
Here, $\mu_N(a)$ represents the $N$-th eigenvalue of the operator $a\in \mathcal{A}$, $m_N$ represents the multiplicity of this eigenvalue, and $D^2$ is the number operator of the harmonic oscillator system itself. An explanation of the geometric meaning of $D$ is in order. If $\mu$ and $\nu$ are the Euclidean space-time indices, then the relation $D\! =\! \frac{1}{\sqrt{2}}\gamma^{\mu}\hat{p}_{\mu}$ follows from the relation $D^2\! =\! \frac{1}{2}\hat{p}_{\mu}^2$, and this surely implies that it is the extension of the Dirac operator $D\! =\! \gamma^{\mu}\partial_{\mu}$ on the ordinary commutative space. On the other hand, if $\mu$ and $\nu$ are the Minkowskian space-time indices, then $D^2$ and $\frac{1}{2}\hat{p}_{\mu}\hat{p}^{\mu}$ have different signatures. It is not yet clear whether choosing a representation other than (\ref{eq:11}) allows us to obtain a $D^2$ that reproduces the Minkowskian signature, or if the Euclidean metric necessarily emerges in the infinitesimal region in the noncommutative space.

Next, we introduce the Dixmier trace $\tr_{\omega}$ for the operators of the harmonic oscillator, which is written in terms of the infinite-dimensional matrices\cite{rf:landi} as 
\begin{align}
	\tr_{\omega}a\equiv \lim_{N\to \infty}
	\frac{\displaystyle \> \sum_{n=0}^N m_n\mu_n(a)\>}
	{\displaystyle \log \sum_{n=0}^N m_n}.
	\label{eq:14}
\end{align}
This trace picks out the coefficients of the logarithmic divergences. For example, the trace of the $(-4)$-th power of the Dirac operator is 
\begin{align}
	\tr_{\omega}D^{-4} & =\lim_{N\to \infty}
	\frac{\displaystyle \> \sum_{n=0}^{N}m_n\mu_n(D^{-4})\>}
	{\displaystyle \log \sum_{n=0}^N m_n}
	=\lim_{N\to \infty}\frac{\theta^2\log N}{2\log N}=\frac{\theta^2}{2},
	\label{eq:15}
\end{align}
and, in general, we have 
\begin{align*}
	\tr_{\omega}D^{s} & =\left\{ \begin{array}{ll}
	\infty , & \text{for }s>-4, \\
	0, & \text{for }s<-4.
	\end{array} \right.
\end{align*}
Utilizing this trace $\tr_{\omega}$, we define the integral $\int$ as follows: 
\begin{align}
	& \int a \equiv \frac{2}{\theta^2}\tr_{\omega}D^{-4}a, \quad 
	a\in \mathcal{A}, 
	\label{eq:16}
\end{align}
with the normalization 
\begin{align}
	& \quad \int 1=1. 
	\label{eq:16.5}
\end{align}
Here, $D^{-4}$ is inserted as the ``integration measure", and, it represents the volume element of the noncommutative space. This identification is consistent with dimensional analysis, because we have $D\! \sim \! \hat{p}_{\mu}$.

Next, let us construct the complete system to expand fields, such as the gauge field. It is sufficient, in principle, to do this for the 1-dimensional harmonic oscillator, and we focus on the generators $(\hat{x}^0,\hat{x}^1)$ and the annihilation and creation operators $(a_1,a_1^{\dagger})$. Here, we first make a conjecture for a complete system, and prove that in fact this is a complete system. The function $a(\hat{x}^0,\hat{x}^1)$ is expanded in terms of ``plane waves" as 
\begin{align}
	a(\hat{x}^0,\hat{x}^1)
	=\sum_{k=\{k_0,k_1\}}a_ke^{ik_0\hat{x}^0+ik_1\hat{x}^1}.
	\label{eq:17}
\end{align}
In order to prove that these ``plane waves" form a complete system, we need to show the relation 
\begin{align}
	& \int f^{\dagger}(k_0,k_1)\cdot f(l_0,l_1)
	=\delta_{k_0l_0}\delta_{k_1l_1}.
	\label{eq:18}
\end{align}
where $f(k_0,k_1)\equiv e^{ik_0\hat{x}^0+ik_1\hat{x}^1}$. This is trivial for the case $k_0\! =\! l_0$ and $k_1\! =\! l_1$ from the relation $\int 1=1$. Otherwise, we need some calculations. As preparation, we find that the elements of the matrices $e^{ika_1}$ and $e^{ika_1^{\dagger}}$ are 
\begin{align}
	& \langle n|e^{ika_1}|m\rangle 
	=\sum_{r}\frac{(ik)^r}{r!}\langle n|a_1^r|m\rangle
	=\frac{(ik)^{m-n}}{(m-n)!}\sqrt{\frac{m!}{n!}}, \nonumber \\
	& \langle n|e^{ika_1^{\dagger}}|m\rangle 
	=\sum_{r}\frac{(ik)^r}{r!}\langle n|a_1^{\dagger r}|m\rangle
	=\frac{(ik)^{n-m}}{(n-m)!}\sqrt{\frac{n!}{m!}}, 
	\label{eq:19}
\end{align}
and therefore the diagonal components of $f^{\dagger}(k_1,k_2)\cdot f(l_1,l_2)$ are 
\begin{align}
	\langle n|f^{\dagger} & (k_0,k_1)\cdot f(l_0,l_1)|n\rangle \nonumber \\
	& =e^{-\frac{\theta}{2}(l_0-k_0)^2-\frac{\theta}{2}(l_1-k_1)^2
	+i(k_0l_1-k_1l_0)\theta} \nonumber \\
	& \qquad \quad \times \sum_{m=0}^n \frac{(-1)^m
	\left\{\frac{\theta}{2}(l_0-k_0)^2+\frac{\theta}{2}(l_1-k_1)^2\right\}
	^m}{m!}\frac{n!}{m!(n-m)!} \nonumber \\
	& \equiv e^{-\frac{\theta}{4}(l_0-k_0)^2-\frac{\theta}{4}(l_1-k_1)^2+
	i(k_0l_1-k_1l_0)\theta}F([-n],[1],
	\textstyle \frac{\theta}{2}(l_0-k_0)^2+\frac{\theta}{2}(l_1-k_1)^2).
	\label{eq:20}
\end{align}
\begin{figure}[htb]
	\centerline{\includegraphics*[width=10.5cm,angle=0]{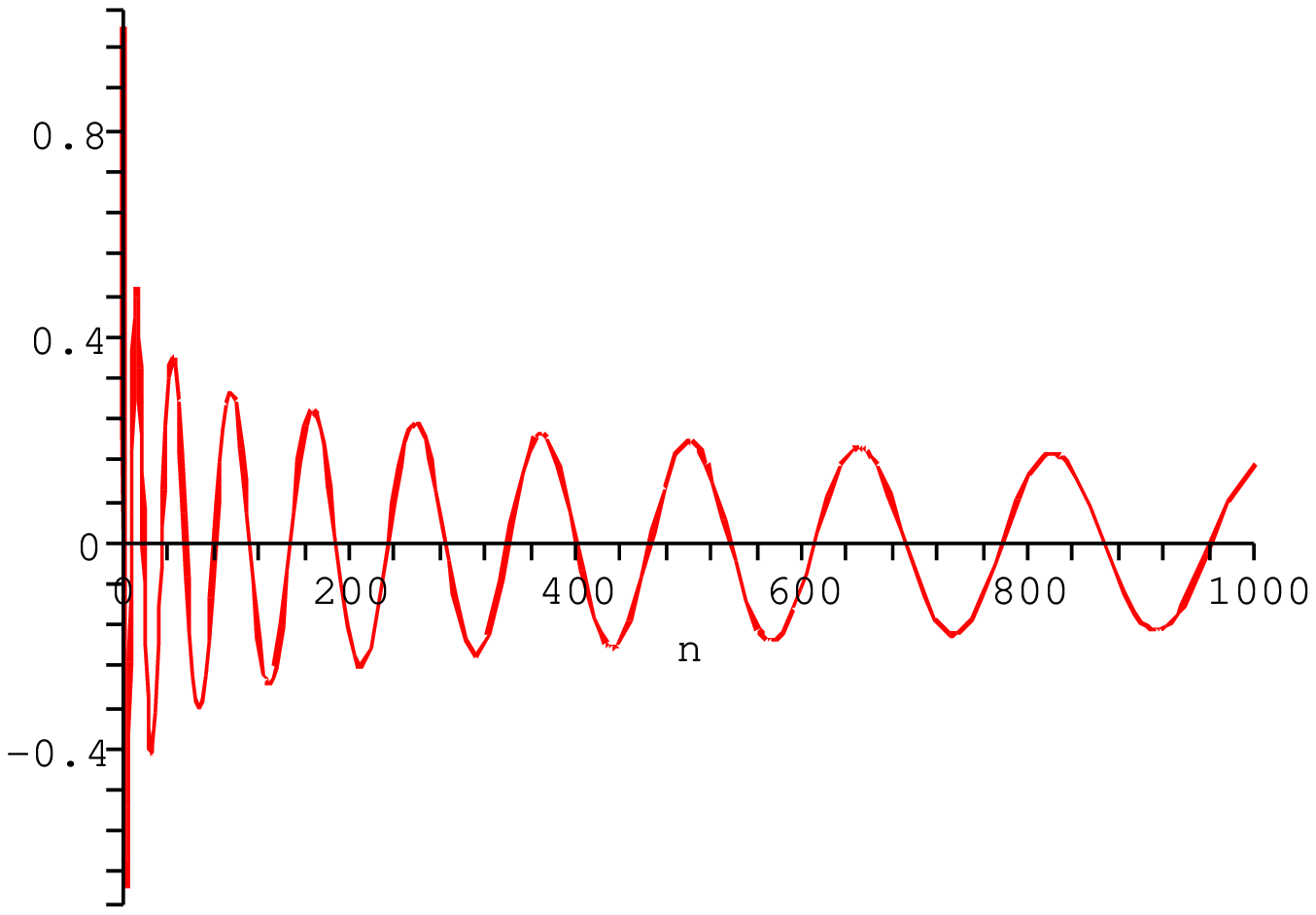}}
	\caption{$F=F([-n],[1],z=1)$}
	\label{fig:1}
	\centerline{\includegraphics*[width=10.5cm,angle=0]{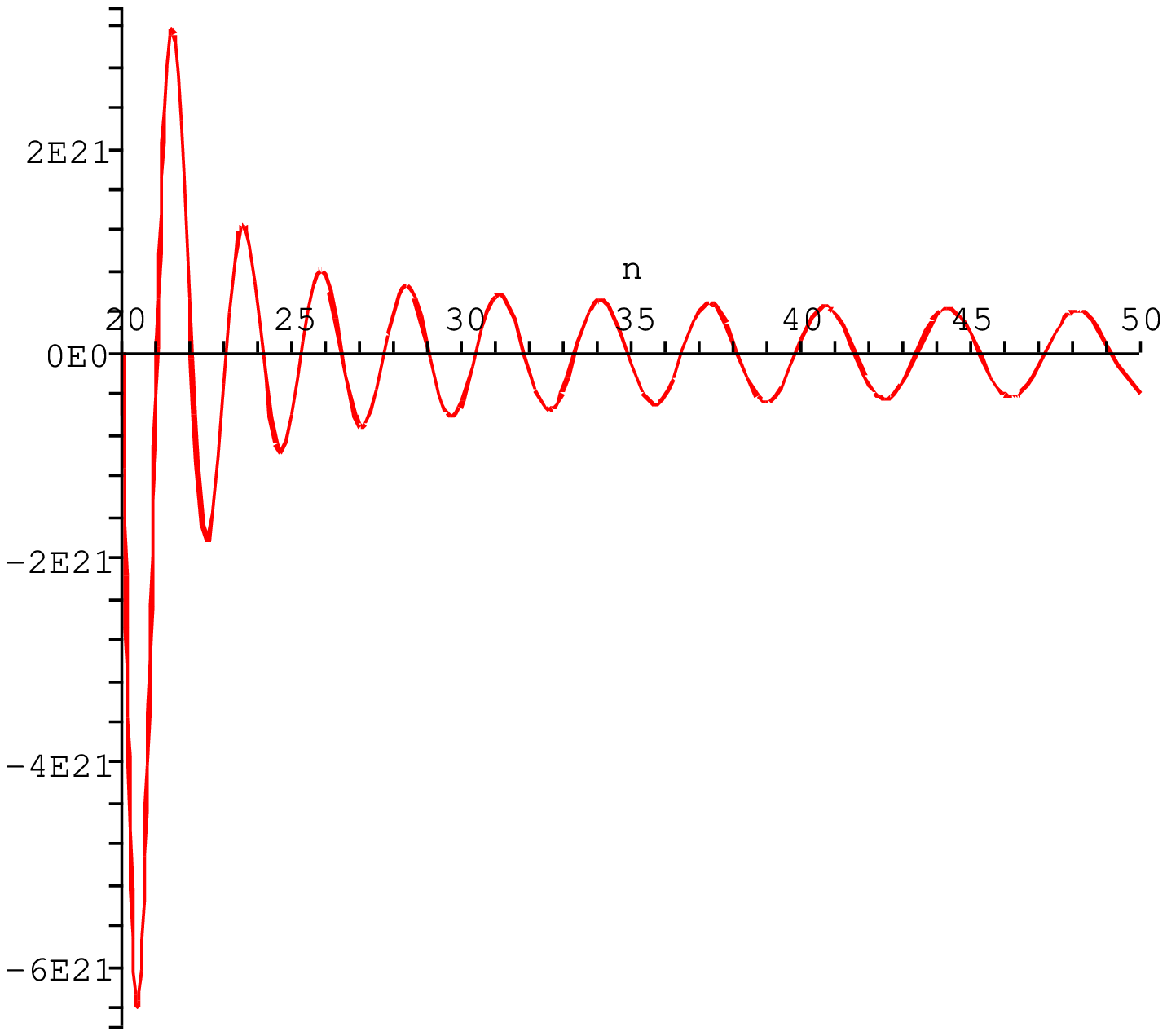}}
	\caption{$F=F([-n],[1],z=100)$}
	\label{fig:2}
\end{figure}
The quantity $F([-n],[1],z)\! =\! \sum_{m=0}^n\frac{(-1)^mz^m}{m!}\frac{n!}{m!(n-m)!}$ is a Laguerre polynomial and, as shown in Figs. 1 and 2, it decays to zero as $n\! \to \! \infty$ for fixed $z$. As $z\! =\! \frac{\theta}{2}(l_0-k_0)^2+\frac{\theta}{2}(l_1-k_1)^2$ increases, the oscillation frequency of $F([-n],[1],z)$ increases, and the rate at which it damps to zero also increases. Therefore (\ref{eq:18}) is satisfied. Furthermore, we find that the integral $\int$ is an operation that draws out the zero mode: 
\begin{align}
	\int a(\hat{x})=\int \sum_{k=\{k_{\mu}\}}a_ke^{ik_{\mu}\hat{x}^{\mu}}
	=a_{k=0}.
	\label{eq:21}
\end{align}

Now we must show that the action (\ref{eq:10}) is invariant under the gauge transformation (\ref{eq:7}). To do this, it is sufficient to demonstrate the relation 
\begin{align}
	\int U(\hat{x})a(\hat{x})=\int a(\hat{x})U(\hat{x}), \qquad 
	a(\hat{x})\in \mathcal{A} 
	\label{eq:22}
\end{align}
for a unitary operator $U(\hat{x})$. Because $a(\hat{x})$ and $U(\hat{x})$ can be expanded as 
\begin{align}
	a(\hat{x}) & =\sum_{k=\{k_{\mu}\}}a_ke^{ik_{\mu}\hat{x}^{\mu}}, 
	\nonumber \\
	U(\hat{x}) & =e^{iH(\hat{x})}, \qquad 
	H(\hat{x})\equiv \sum_{k=\{k_{\mu}\}}H_ke^{ik_{\mu}\hat{x}^{\mu}} 
	\nonumber \\
	& \equiv \sum_{l=\{l_{\mu}\}}U_le^{il_{\mu}\hat{x}^{\mu}}, 
	\label{eq:23}
\end{align}
(where $H_k$ is hermitean, i.e. $H_k^{\ast}=H_{-k}$) in the plane waves, we only need to show the identity 
\begin{align}
	\int [e^{ik_{\mu}\hat{x}^{\mu}},e^{il_{\mu}\hat{x}^{\mu}}]=0, 
	\label{eq:24}
\end{align}
and this is obvious from the above proof of the completeness (\ref{eq:18}). We thus find that the action (\ref{eq:10}) is invariant under the gauge transformation (\ref{eq:7}).

Next, note that $End_{\mathcal{A}}\mathcal{H}$ generated from $\hat{x}^{\prime \mu}$ has the exactly same structure as $\mathcal{A}$, except for the signature of $\theta^{\mu \nu}$, as shown in (\ref{eq:1}), and therefore the definition of the integral $\int$ and the gauge invariance can be obtained just as in the case of $\mathcal{A}$.

Finally, let us evaluate the integral $\int$ of the action (\ref{eq:10}). In the algebras $\mathcal{A}$ and $End_{\mathcal{A}}\mathcal{H}$, the field strengths are 
\begin{align}
	& A_{\mu}(\hat{x})
	=\sum_{k=\{k_{\mu}\}}A_{\mu ,k}e^{ik_{\mu}\hat{x}^{\mu}}, 
	\quad A_{\mu ,k}^{\ast}=A_{\mu ,-k}, \nonumber \\
	& [A_{\mu}(\hat{x}),A_{\nu}(\hat{x})]
	=\sum_{k,l}A_{\mu ,k}A_{\nu ,l}\> e^{i(k+l)_{\mu}\hat{x}^{\mu}}
	(-2i)\sin \frac{k_{\mu}l_{\nu}\theta^{\mu \nu}}{2} 
	\label{eq:25}
\end{align}
and 
\begin{align}
	& A_{\mu}(\hat{x}^{\prime})
	=\sum_{k=\{k_{\mu}\}}A_{\mu ,k}
	e^{ik_{\mu}\hat{x}^{\prime \mu}}, 
	\quad A_{\mu ,k}^{\ast}=A_{\mu ,-k}, \nonumber \\
	& [A_{\mu}(\hat{x}^{\prime}),A_{\nu}(\hat{x}^{\prime})]
	=\sum_{k,l}A_{\mu ,k}A_{\nu ,l}\> e^{i(k+l)_{\mu}\hat{x}^{\prime \mu}}
	(+2i)\sin \frac{k_{\mu}l_{\nu}\theta^{\mu \nu}}{2}, 
	\label{eq:26}
\end{align}
respectively. The difference between the signatures in front of $\sin \frac{k_{\mu}l_{\nu}\theta^{\mu \nu}}{2}$ comes from the difference between the signatures of $\theta^{\mu \nu}$ for $\mathcal{A}$ and $End_{\mathcal{A}}\mathcal{H}$. Then, we find that the Yang-Mills action $S_{\text{YM}}$ of (\ref{eq:10}) is 
\begin{align}
	S_{\text{YM}}
	& =-2\sum_{k,l,m,n}A_{\mu ,k}A_{\nu ,l}A^{\mu}_mA^{\nu}_n
	\sin \frac{k_{\mu}l_{\nu}\theta^{\mu \nu}}{2}
	\sin \frac{m_{\mu}n_{\nu}\theta^{\mu \nu}}{2} \nonumber \\
	& \qquad \qquad \qquad \qquad \times 
	\cos \frac{(k+l)_{\mu}(m+n)_{\nu}\theta^{\mu \nu}}{2} 
	\delta_{k+l+m+n,0} 
	-\frac{1}{2}\theta_{\mu \nu}^{-1}\theta^{-1 \mu \nu}. 
	\label{eq:27}
\end{align}
We can confirm from the nature of the trigonometric function that the exchange of $k$ with $l$ or $m$ with $n$ alters the sign of $S_{\text{YM}}$, while that of the set $\{k,l\}$ with $\{m,n\}$ does not. This reflects the index structure of the square of the field strength, $F_{\mu \nu}F^{\mu \nu}$.

From the above results, although it is naive, we can conclude that the definition of the quantization using the path integral formalism is 
\begin{align}
	& Z_M(\theta )=\int \prod_{\mu}\prod_{k=\{k_{\mu}\}}dA_{\mu ,k}\> 
	e^{iS_{\text{YM}}} \quad \text{for the Minkowskian case,} \nonumber \\
	& Z_E(\theta )=\int \prod_{\mu}\prod_{k=\{k_{\mu}\}}dA_{\mu ,k}\> 
	e^{-S_{\text{YM}}} \quad \text{for the Euclidean case.}
	\label{eq:28}
\end{align}
In order to carry out the quantization exactly, we must investigate the gauge fixing, the treatment of the ghosts and more. Such treatments are the topics of future works. Here we take $k$ to be an integer in the coefficients $A_{\mu ,k}$ of the expansion of the field $A_{\mu}(\hat{x})$. As a result, this system is invariant under the shift $\hat{x}^{\mu}\! \to \! \hat{x}^{\mu}+2\pi$. That is to say, we should regard the noncommutative space on which our gauge theory is defined as noncommutative torus. Then, if we want to define the gauge theory on the noncommutative Minkowskian (or Euclidean) flat space, we can change $k$ to a real number, employing an almost straightforward procedure. The extension to the supersymmtric case is also made trivial by, for example, utilizing the superspace formalism of the noncommutative space presented in Ref.~\citen{rf:y2}.

\section{Conclusion}

The purpose of this paper is to present the tools necessary to carry out the quantization of the gauge theory on the canonical noncommutative space. We introduce the Dirac operator that defines the line element of the noncommutative space and define the trace as the volume integration for infinite-dimensional matrices, which represent the algebra of the noncommutative space. When the space-time dimensionality is even (i.e. $d=2r$), we can exactly evaluate the action by utilizing the basis of the eigenfunctions of the $r$-dimensional harmonic oscillator. Then, the spectrum of the coordinate $\hat{x}^{\mu}$ is given by $(\hat{x}^{\mu})^2\! \sim \! \theta^2D^2\! \sim \! \theta (N\! +\! 1)$, from (\ref{eq:13}). This is identically the number operator and is discretized with the width $\sqrt{\theta}$. From this result, we can conclude that the noncommutativity of the space-time certainly removes the infinitesimal space-time region, i.e. the ultraviolet region, and hence we have $[\hat{x}^{\mu},\hat{x}^{\nu}]\! \sim \! i\theta \Rightarrow \Delta x^{\mu}\Delta x^{\nu}\! \geq \! \theta$. Also, considering the procedure of our construction, it is natural to conjecture that the gauge theory on the canonical noncommutative space could reduce to the ordinary gauge theory on the commutative space in the limit $\theta \! \to \! 0$, i.e. the limit in which the ultraviolet cutoff is not incorporated as a precondition of the theory, in contrast to the situation for the noncommutative space. This leads us to propose the following resolution of the cosmological constant problem. The contribution of the vacuum (Casimir) energy may come from the first and second terms of the action $S_{\text{YM}}$ in (\ref{eq:10}), just as in the ordinary gauge theory on commutative space, and it is of the order of $m_p^4$ (which can also be obtained from dimensional analysis). This is $10^{120}$ times larger than the observed cosmological constant $\Lambda$, and thus alone it cannot be regarded as representing $\Lambda$. However, our gauge theory on noncommutative space also has the geometric contribution $-\theta_{\mu \nu}^{-2}$, which comes from the space-time noncommutativity, as shown in the third term of (\ref{eq:10}), and if the space-time noncommutativity emerges near the Planck scale, then we have $\theta^{-2}\! \sim \! m_p^4$. Therefore, it may be plausible that the contributions from the vacuum energy of the gauge field and from the space-time noncommutativity, which has a geometric origin, nearly cancel and thus yield a very small cosmological constant. We will determine rigorously whether or not this mechanism actually does exist in a forthcoming article.

Finally, let us comment on a gravity theory of the noncommutative space. As stated at the end of \S 2 in connection with the fact that the gauge transformation (\ref{eq:7}) includes the geometric one, it may be the case that the action $S_{\text{YM}}$ of (\ref{eq:10}) describes gravity in some sense. In fact, the square of the field strength $F_{\mu \nu}F_{\rho \lambda}\! =\! [A_{\mu},A_{\nu}][A_{\rho},A_{\lambda}]$ has the same space-time index structure as the Riemann tensor $R_{\mu \nu \rho \lambda}$.\cite{rf:kawai} However, provided that it contains gravity, it seems appropriate, within our formalism, to consider this to be a special solution of the Riemann tensor. Therefore, as a future work, we would like to introduce a space-time metric $g_{\mu \nu}(\hat{x})$ and a Dirac operator $D\! \sim \! \gamma^{\mu}\hat{p}^{\nu}g_{\mu \nu}(\hat{x})$ that represents an extension of our ``flat" case (\ref{eq:13}) and defines a line element on a ``curved" noncommutative space.


\end{document}